\documentclass[preprint2]{aastex}

\slugcomment{}

\shortauthors{Olsen et al.}
\shorttitle{}

\begin{document}


\title{A Comprehensive Look at LH72 in the Context of Supergiant Shell LMC-4}


\author{Knut A.G. Olsen}
\affil{Cerro Tololo Inter-American Observatory, Casilla 603, La Serena, Chile}
\email{kolsen@noao.edu}

\author{Sungeun Kim}
\affil{Harvard-Smithsonian Center for Astrophysics, 60 Garden Street, Cambridge, MA 02138}
\email{skim@cfa.harvard.edu}

\and

\author{Jeremy F. Buss}
\affil{University of Wisconsin, Oshkosh, 4930 Schuh Rd., Appleton, WI 54915}
\email{jeremy@sunspot.phys.uwosh.edu}



\begin{abstract}
Stellar spectroscopy, $UBV$ photometry, H$\alpha$ imaging, and analysis of data from the ATCA \ion{H}{1} survey of the LMC are combined in a study of the LMC OB association LH 72 and its surroundings.  LH 72 lies on the rim of a previously identified \ion{H}{1} shell, SGS-14, and in the interior of LMC-4, one of the LMC's largest known supergiant shells.  Our analysis of the \ion{H}{1} data finds that SGS-14 is expanding with velocity $v_{exp}\sim15$ km s$^{-1}$, giving it an expansion age of $\sim$15 Myr.  Through the stellar spectroscopy and photometry, we find similar ages for the oldest stars of LH 72, $\sim$15$-$30 Myr.  We confirm that LH 72 contains an age spread of $\sim$15$-$30 Myr, similar to the range in ages of stars derived for the entire surrounding supergiant shell.  Combining analysis of the O and B stars with H$\alpha$ imaging of the \ion{H}{2} region DEM 228, we find that DEM 228 accounts for only 60\% of the available ionizing Lyman continuum photons.  Comparing the distribution of ionized gas with that of the \ion{H}{1}, we find that DEM 228 and LH 72 are offset by $\sim1-2\arcmin$ from the peak 21-cm emission, towards the interior of SGS-14.  Taken together, these results imply that SGS-14 has cleared its interior of gas and triggered the formation of LH 72.  On the basis of our results, we suggest that LMC-4 was not formed as unit but by overlapping shells such as SGS-14, and that LH 72 will evolve to produce a stellar arc similar to others seen within LMC-4.

\end{abstract}




\section{Introduction}
The Constellation III region in the LMC has drawn considerable interest over the last 50 years, for a number of reasons.  One is that the region contains several structures noteworthy enough to be given names.  At the wavelength of H$\alpha$, the most striking feature is a kpc-sized ring of \ion{H}{2} regions, dubbed LMC-4 by \citet{m80}.  The void of LMC-4's interior is almost empty of other ionized gas, except for a single bright \ion{H}{2} region, the ``eighth-note nebula'' DEM 228 \citep{dem}.  21-cm maps, such as the high-resolution map of \citet{k98}, show that LMC-4 is enclosed by a near-circular ring of \ion{H}{1} gas, named SGS-11 by \citet{k99}.  The ring shape of \ion{H}{1} is only spoiled by a tongue of gas that joins DEM 228.  \citet{lh70}, in their mapping and analysis of OB associations in the LMC, identified many associations contained within the boundaries LMC-4.  Some of these are organized into circular arcs of stars, two of the most prominent of which were named ``Quadrant'' and ``Sextant'' by \citet{ee98}.  The colorful name Constellation III, as originally used by \citet{ms53}, designates the Sextant arc \citep{ee98}.

More important than the names are the variety of physical mechanisms suggested to have created the kpc-sized hole in the interstellar medium and the substructures within (see review in \citet{ee98}).  The size of the hole suggests a large input of energy required to clear it, perhaps produced in a ``super-supernova'' \citep{wm66}, a gamma ray burst \citep{eeh}, or a high-velocity cloud impact with the LMC disk \citep{b96}.  \citet{dmf} offered a less dramatic solution, in which star formation has propagated outwards from a central location, clearing a cavity through the combined action of supernovae and stellar winds.  The predicted age gradient in the stellar population interior to LMC-4 has not been observed in the most recent work \citep{o97, b97, dh98}; however, stochastic self-propagating star formation (SSPSF) is often still believed to play some role in the formation of LMC-4.  An age gradient would not be produced if star formation within LMC-4 initiated over a large area at the same time, as in the bow shock-induced star formation scenario of \citet{db98}.

The models above focus mostly on the size of LMC-4's central cavity, and less so on the substructure of the interior stellar population.  \citet{ee98} tackled the problem of the substructure by considering the formation of the Quadrant and Sextant stellar arcs.  They found evidence for a concentration of 30 Myr-old stars close to the center of the Quadrant arc; if these stars represent the remnants of a stellar cluster, they could provide the necessary pressure to sweep up a shell of \ion{H}{1} gas out of which Quadrant could have condensed.  \citet{ee98} thus showed that smaller-scale star formation events have contributed, in some degree, to the formation of LMC-4.

For a comprehensive understanding of how LMC-4 formed and has evolved, with attention paid to both the size of the supershell and its substructure,
it will be necessary to survey its entire massive star population, measure the properties of all of its individual \ion{H}{2} regions, and study the relationship of the stars and ionized gas with the full distribution and kinematics of the neutral atomic and molecular gas.  While many of these pieces are coming into or are already in place, much remains to be done, particularly in the areas of stellar spectroscopy and the analysis of large existing datasets.  This work takes a small step in the direction of taking a comprehensive look at LMC-4 by focussing on the individual OB association LH 72, its embedded \ion{H}{2} region DEM 228, and its associated \ion{H}{1} gas.  Because of its isolation, LH 72 is a relatively uncomplicated object with which to begin the analysis.  In addition, as remarked by \citet{o97}, its location in LMC-4's interior may yield a special clue as to the formation and evolution of the surrounding supershell.  Figure \ref{fig0} displays images of LH 72 and DEM 228, showing their relationship to SGS-14 and LMC-4.

\section{Observations}
Optical imaging and spectroscopy of LH 72 and narrow-band imaging of the embedded \ion{H}{2} region DEM 228 were performed during three runs with the CTIO 1.5-m and 0.9-m telescopes.  The primary broad- and narrow-band imaging was carried out on the nights of November 24-26, 1998 with the 1.5-m Cassegrain Focus imager at f/7.5 using the SITe 2048$\times$2048 CCD \#6.  The f/7.5 secondary mirror produces a pixel scale of 0\farcs44 pixel$^{-1}$, yielding a 15\arcmin$\times$15\arcmin ~field of view at the expense of a somewhat erratically varying PSF.  We observed through 3$\times$3-inch Johnson $UBV$ filters and 2$\times$2-inch narrow-band H$\alpha$ and continuum filters.  The 2-inch filters vignette the edges of the field, leaving a $\sim$10\arcmin$\times$10\arcmin ~useful area.  Total object exposure times were 1680 sec in $U$, 690 sec in $B$, and 600 sec in $V$.
We also observed \citet{l92} and \citet{g82} standards at a variety of airmasses each night so as to calibrate the $UBV$ photometry.  Calibration of the H$\alpha$ images was achieved through 1.5-m and 0.9-m observations of spectrophotometric standards from \citet{h94} and selected \ion{H}{2} regions in the LMC and NGC 6822.

Spectra of 12 of the brightest blue stars in LH 72 were taken with the 1.5-m R-C spectrograph and Loral 3K detector on the nights of December 30, 1998 and January 2-5, 1999.  Using grating \#58 in second order and a CuSO$_4$ filter to block contamination from first order, our spectra cover the wavelength range 3600-5000\AA, spanning the classical spectral classification region of 4000$-$4700\AA.  Exposure times were chosen to achieve S/N$\sim75-100$ per resolution element.  Comparison lamp exposures were taken at regular intervals to provide wavelength calibration.  Observations of the spectrophotometric standards EG 21 and Hiltner 600 \citep{h94} were used to provide flux calibration.

\section{The HI data}
The \ion{H}{1} data analyzed here comes from a combination of the ATCA 21-cm survey of the Magellanic Clouds \citep{k98} and Parkes single-dish observations of the LMC.  The mosaic of interferometer data gives a spatial resolution of 1\arcmin ~and velocity resolution of 1.6 km s$^{-1}$.  The single dish observations provide for the zero-spacing flux, so that the total mass in \ion{H}{1} may be accurately calculated.  For this project, we worked on a subset of the \ion{H}{1} data cube covering a 48\arcmin$\times$48\arcmin ~region centered on 5$^{\rm h}$23$^{\rm m}$34\fs7 and -69\arcdeg44\arcmin22\arcsec (2000.0) and spanning heliocentric velocities between 190 and 390 km s$^{-1}$.

\section{Basic reduction procedures}
All of the CTIO frames were overscan-corrected, bias-subtracted, and flat-fielded using IRAF routines with calibration data taken at the beginning and end of each night.  For the imaging program, we used twilight sky flats to correct for pixel-to-pixel sensitivity variation.  Because many of the flat-field and standard star exposure times were necessarily short, we applied a shutter shading correction to images with exposure times less than 20 seconds.  The shutter shading map was derived by comparing 20-second $R$ dome flats with $R$ dome exposures during which the shutter was repeatedly opened for one second and then closed.  We also produced a bad pixel mask from short- and long-exposure $R$ domes, which we applied to the final reduced images.

We discovered during the later analysis that all of the 1.5-m CCD images had been written to the data acquisition computer missing the central row of pixels.  The problem manifested itself as a 1-pixel shift in the positions of stars along the $y$-axis between dithered exposures.  To account for the dropped row, we manually inserted a row of pixels into all of the images, while flagging the row in the bad pixel mask.

For the spectral program, we used quartz lamp dome flats to correct for pixel sensitivity.  Before applying the flats to the data, we corrected them for the shape of the quartz lamp spectrum and for the slit illumination function by fitting the flats with a cubic spline surface.  Obvious cosmic rays were manually removed from the flat-fielded data by linearly interpolating over the bad pixels along the spatial direction.  Following experimentation with its parameters, we used the IRAF package {\it doslit} to extract wavelength- and flux-calibrated spectra.  While flux calibration may seem unnecessary for spectral classification work, we found it easier to fit the continuum after the flux calibration was applied, as the spectral response curve is wiggly compared to the smooth continuum of the tail of the blackbody spectrum. 
Our final spectral analysis was performed on continuum flattened spectra, boxcar smoothed with a 3-pixel-wide kernel.  For the stars with multiple observations, we combined the continuum-flattened extracted spectra before smoothing.

\section{Analysis of the stellar content}
\subsection{Spectral classification}
The stellar spectra were classified independently by two of the authors (KO and JB).  Following the criteria of \citet{wf90}, we identified first the spectral type and next the luminosity class.  Briefly, for the O stars observed here, comparison of the strengths of HeII 4541\AA ~to HeI 4471\AA ~and HeII 4200\AA ~to HeI+II 4026\AA ~provide the temperature; the strengths of SiIV 4089\AA ~and HeII 4686\AA ~are the main luminosity indicators.  For the early B stars, the temperature decision depends mainly on the relative strengths of SiII 4552\AA ~and SiIV 4089\AA, while luminosity is decided through SiIV 4089\AA, SiIII 4552\AA, or the strength of MgII 4480\AA.  In all cases, our identifications agreed to within two, and usually one, spectral subtypes and one luminosity class.

Figures \ref{fig1}a\&b show the continuum-subtracted spectra of the 12 stars along with identified lines.  Three of the stars, L10, L39, and S132, show narrow emission lines indicative of circumstellar disks.  Table \ref{tbl-1} lists our identifications.  Where appropriate, the spectral type represents the average of our independent judgings.

\subsection{Stellar photometry}
$UBV$ photometry of LH 72 was performed with the DAOPHOT and ALLSTAR suite of programs \citep{s87}.  Aperture and point spread function (PSF)-fitting photometry was performed on individual frames, with results from multiple frames averaged at the end.  We first performed aperture photometry, using a 7\arcsec ~radius for the largest aperture and 1\farcs75 for the smallest.  Using DAOGROW, we produced a growth curve to adjust the magnitudes measured within the smaller apertures so as to agree with the largest.  DAOGROW produced a single list of magnitudes for each star, using the largest possible aperture with photometric error $\sigma<0.02$ magnitudes, with the magnitude extrapolated to an aperture of 14\arcsec ~radius.

We next performed PSF photometry with ALLSTAR.  Approximately 40 PSF stars were chosen automatically by DAOPHOT, the list of which was then edited by hand to select only uncrowded stars.  A PSF was derived for each image by fitting a Moffat function to the stars and employing a quadratically varying empirical lookup table to correct for deviations from the analytic function.  Aperture corrections were derived from the previously measured aperture magnitudes, extrapolated to 14\arcsec ~radius.  Finally, we merged all of the aperture and PSF magnitudes through Stetson's COLLECT, which produced a weighted average magnitude for each star in each filter, using the PSF magnitudes whenever those photometric errors were found to be smaller.

The instrumental magnitudes were placed on the standard $UBV$ system through photometry of \citet{l92} standard fields SA 92, SA 98, SA 101, and T Phe and Graham's (1982) E2 and E3 fields.  The standard star photometry was performed as for the object frames, including PSF photometry, as the faintest stars were underexposed.  We derived transformations through least-squares fitting to the equations:
\begin{displaymath}
u = U + A_0 + A_1(U-B) + A_2(B-V)^2 + A_3X + A_4t
\end{displaymath}
\begin{displaymath}
b = B + B_0 + B_1(B-V) + B_3X + B_4t
\end{displaymath}
\begin{displaymath}
v = V + C_0 + C_1(B-V) + C_3X + C_4t
\end{displaymath}
where lower-case letters represent instrumental magnitudes, upper-case letters standard magnitudes, $X$ is the airmass, and $t$ the time of observation in hours from the middle of the night.  The second-order color term $A_2$ for $U$ is necessary because of the steep response curve of the CCD in the ultraviolet.
While usually small, variation of the zero point with time may make a difference at the level of a few hundredths of a magnitude.

After deriving the color coefficients $A_1, A_2, B_1,$ and $C_1$ independently for each night, we took their average and used them to rederive the remaining nightly coefficients.  We found the average scatter around the transformation relations to be 0\fm035 in $U$, 0\fm015 in $B$, and 0\fm008 in $V$.
Table \ref{tbl-2} lists the coefficients for the three photometric nights.

To place the LH 72 photometry on the $UBV$ system, we first transformed the photometry derived from frames taken on 11/25/98 using the relation computed for that night.  We then selected 25 isolated stars from the LH 72 frame and used their computed $UBV$ magnitudes to transform the photometry taken on the one non-photometric night.  Finally, all of the standardized magnitudes were averaged to produce a master photometric table for LH 72.

Figure \ref{fig2} shows the comparison of our photometry with that from the Magellanic Clouds Photometric Survey (MCPS; Zaritsky, Harris, \& Thompson 1997), kindly provided by Dennis Zaritsky.  Our photometry agrees within 3$\sigma$ of that of the MCPS: $U-U_{\rm MCPS} = 0.005\pm0.010, B-B_{\rm MCPS} = 0.017\pm0.009, {\rm and} V-V_{\rm MCPS} = 0.033\pm0.009$.  We also find good agreement with the photographic photometry of \citet{l72}: $V-V_{\rm L72} = 0.022\pm0.007$ (39 stars), $(B-V)-(B-V)_{\rm L72} = 0.01\pm0.02$ (39 stars), and $(U-B)-(U-B)_{\rm L72} = -0.05\pm0.03$ (5 stars).

An additional check on the photometry is to compare the intrinsic $(U-B)_\circ$ color expected from the spectral classification to that derived from the photometry after correcting for reddening.  Figure \ref{fig3} shows the comparison; we find $(U-B)_{\rm \circ,spec}-(U-B)_{\rm \circ,phot} = -0.03\pm0.01$, suggesting good agreement between the photometry and spectroscopy.  There is, however, a possible discrepancy at the bluest and reddest ends of the scale.

\subsection{Analysis of stellar photometry}

\subsubsection{LH 72 membership}
Figure \ref{fig4}a shows the raw $B-V,V$ color-magnitude diagram (CMD) produced from our photometry.  As is typical of ground-based composite LMC CMDs, we see a broad main sequence down to $V\sim21$, an extended giant branch, and a prominent red clump.  The presence of an OB association is clearly evidenced by the well-populated upper main sequence; however, the membership of LH 72 is confused by the background field population.  To restrict our analysis to probable association members, we selected stars lying within the boundary defined by \citet{l72}.   Defining the remaining stars as representatives of the background population, we subtracted the background CMD from the CMD of stars lying within LH 72's boundary, as done e.g. by \citet{o99}.  Figure \ref{fig4}b shows the result of the subtraction procedure.  We see that most of the lower main sequence, giant branch, and red clump have disappeared, while most of the upper main sequence remains.  At $V<16$, $\gtrsim$80\% of the stars are probable members of LH 72; between $16<V<17$, the contamination rate is $\sim$50\%.  For the purposes of our analysis, therefore, we choose as LH 72 members only those stars with $V<16$ in the background-subtracted CMD.

\subsubsection{Identification of Be stars}
As discussed by e.g. \citet{mgm99}, the presence of emission lines in the spectra of late O and early B stars (the Be star phenomenon) is indicative of circumstellar disks.  Fast rotation and the presence of circumstellar material causes Be stars to generally be redder and cooler than their non-Be counterparts, a fact which unaccounted for may be misinterpreted as an age difference.

As done by e.g. \citet{g97}, we identified Be star candidates photometrically by looking for the presence of excess H$\alpha$ emission.  H$\alpha$ magnitudes were measured from the calibrated, continuum-subtracted H$\alpha$ image of DEM 228 (discussed in section 6) through aperture photometry with DAOPHOT.  Figure \ref{fig5} plots the H$\alpha$ 
magnitudes versus $B-V$.  The spectroscopically identified emission-line stars L10 and S132 show an H$\alpha$ excess that separates them from the bulk of the spectroscopically studied stars.  Thus, we label those stars with H$\alpha$ magnitudes $<1.6$ and $B-V<0.5$ as possible Be stars.  Figure \ref{fig6} shows that the stars selected according to this definition are indeed redder than the bulk of the main sequence.  However, we do expect some contamination from non-Be stars, as S128, which does not contain obvious Balmer emission, is included as a Be candidate by our definition.

Because we apply a magnitude cutoff of $V<16$ for certain LH 72 membership, corresponding to $M_V\sim-2.5$ and spectral type $\sim$B1 on the main sequence, we cannot compute the Be star fraction of LH 72 over the full range of possible spectral types.  We note, however, that the fraction is at least 10\% from the number of Be stars found within the boundaries of LH 72.

\subsection{Reddening and the HR diagram}
We computed the reddenings, temperatures, and luminosities of individual stars through the transformations employed by \citet{m00}; these are based on the calibrations of \citet{v96} for the O stars and a combination of sources, including \citet{hm84}, as well as more modern work, for the B stars.  For the stars with spectra, we applied the \citet{v96} and \citet{hm84} calibrations directly; we then used the spectral type-intrinsic color relationship of \citet{f70} and our measured $B-V$ colors to derive $E(B-V)$.

For those stars observed only in $UBV$, we used Massey et al.'s (2000) fits of $T_{\rm eff}$ to the reddening-free index $Q$.  
We picked a luminosity class for each star by plotting Schmidt-Kaler's (1982) magnitude-color relationship on our CMD, assuming $(m-M)_{\rm LH72}=18.4$.  While we found supergiants easy to distinguish from main sequence stars, separating giants from main sequence stars was made difficult by the differential reddening.  Over the range of observed $Q$, confusing a giant with a main sequence star will introduce an error of up to $\sim$0.05 dex in log $T_{\rm eff}$ and $\sim$0.3 magnitudes in $M_{\rm bol}$.
  With $T_{\rm eff}$ and the bolometric correction known, we again used the spectral type-intrinsic color relationship of \citet{f70} to derive $E(B-V)$.

Figure \ref{fig7} shows the $E(B-V)$ distribution for those LH 72 stars with spectra or $Q<-0.4$.  The distribution has a mean of $E(B-V)=0.09$, median of 0.075, standard deviation of 0.055, and a sharp lower bound at $E(B-V)=0.04$.  The typical error on an individual value is $\pm0\fm01$, indicating that the width of the reddening distribution is produced by differential reddening within the LMC and not purely photometric errors.  The typical foreground reddening of $E(B-V)=0.075$ towards the LMC \citep{sfd98} agrees with our median value but not with the minimum.  Outside the boundaries of LH 72, the reddening also has a lower bound $E(B-V)=0.04$, with the majority of stars having $E(B-V)=0.07 \pm 0.04$.

Figure \ref{fig8} shows the combined HR diagram for stars with spectra and stars with photometry only, overlaid with stellar evolutionary tracks and isochrones from \citet{s93}.  With the exception of the O6V star S133, all of LH 72's stars are less massive than 40 $M_\odot$ and appear older than 4$-$5 Myr.  However, as has been seen in a number of LMC OB associations, LH 72 appears to contain a considerable spread in age; the H-R diagram suggests that LH 72 contains stars as old as $\sim$30 Myr.  We see no evidence for the gradient in age that we reported in \citet{o97}; we surmise that the gradient was an artifact of the small number of stars analyzed in that paper.

There are a number of ways besides an age spread within the association to produce a broadened sequence in the observed H-R diagram.  Star-to-star differences in the unknown rotation rates may explain the broadening.  As shown by \citet{hl00}, fast rotation moves main sequence stars to cooler temperatures and lower luminosities, as compared to the non-rotating main sequence.  However, the addition of rotation to a stellar evolutionary track also increases the main sequence lifetime.  The increase in lifetime is such as to produce a real age spread of $\gtrsim 15$ Myr if the broadening in LH 72's H-R diagram were due entirely to stellar rotation, a number very similar to what we claim on the basis of non-rotating stellar models.  A second possibility is that line-of-sight contamination within the LMC produces the broadened sequence.  However, because our analysis shows that 80\% of the OB stars with $V<16$ lie within the confines of LH 72's boundary, we consider this possibility unlikely.

A third, more likely, possibility is that a systematic error in the photometry is responsible for the broadened sequence.  During our comparison of the intrinsic colors of the stars with spectra with their observed $Q$ values, we found good agreement of the colors derived from spectroscopy with those from photometry.  However, if we assume that we should lower the $Q$'s of stars with only photometry by as much as 0\fm05, then we produce the H-R diagram of Figure \ref{fig9}.  The shift in $Q$ makes LH 72 appear younger; the envelope of the youngest stars now suggests an age of $\sim$3 Myr instead of 5 Myr.  The apparent age spread is reduced by a factor $\sim$2; the oldest stars now have ages of $\sim$20 Myr, suggesting an age spread of $\sim$15 Myr.  Such an age spread is still significant compared to the measurement errors in $T_{\rm eff}$ and $M_{\rm bol}$.

\section{Analysis of the ionized gas}
\subsection{Calibration}
The individual H$\alpha$ and off-band images of DEM 228 were first background-subtracted.  We used a 400 pixel-wide box placed in one corner of each image to measure the modal sky value.  The observations of spectrophotometric standards were next used to measure the H$\alpha$ zero point and extinction coefficient.  
To correct for the different illumination of the filter passband by the standards and by the \ion{H}{2} region, we used CTIO's published filter transmission curve and the \citet{h94} published spectra to calculate the number of photons detected at the LMC Doppler-shifted wavelength of H$\alpha$.  The flux of LMC H$\alpha$ photons detected through observation of a standard star is:
\begin{displaymath}
N_{H\alpha} = N \frac{\int F(\lambda)T(\lambda)\delta(\lambda_{H\alpha})d\lambda}{\int F(\lambda)T(\lambda)d\lambda}
\end{displaymath}
where $N$ is the flux of photons observed through the filter, $F(\lambda)$ is the calibrated spectrum of the star, $T(\lambda)$ is the filter transmission curve, and H$\alpha$ has been redshifted to the mean velocity of LH 72's associated \ion{H}{1}, $v=307$ km s$^{-1}$ \citep{k98}.  We used quadratic interpolation to bring Hamuy et al.'s spectra, which are tabulated at 50\AA ~intervals, to the wavelength grid of $T(\lambda)$.  
We then measured the zero point and extinction coefficient by fitting for $Z_{H\alpha}$ and $k_{H\alpha}$ in the equation:
\begin{displaymath}
2.5\log_{10}N_{H\alpha} =  Z_{H\alpha} + k_{H\alpha}X + 2.5\log_{10}{\int F(\lambda)\delta(\lambda_{H\alpha})d\lambda}
\end{displaymath}
where $X$ is the airmass.  We measured $k_{H\alpha}=0.08\pm0.01$, in good agreement with the usual $R$ extinction coefficient of 0.085 at Cerro Tololo. 

After applying the calibration to the individual H$\alpha$ images, we removed cosmic rays and subtracted the off-band images.  Cosmic rays were identified in both H$\alpha$ and off-band images with the IDL Astronomy User's Library routine CR\_REJECT, and removed by linear interpolation through the surrounding good pixels.  We next subtracted the nearest available off-band image, shifting the image to adjust for fractional pixel offsets and scaling it to match the fluxes of point sources through the H$\alpha$ filter.  Finally, we shifted and median combined the calibrated, continuum-subtracted H$\alpha$ images into a single image.

\subsection{Correction of H$\alpha$ flux for reddening}
As shown by the $UBV$ photometry, variable reddening across DEM 228 is significant.  One way to correct for the spatially varying reddening would be to observe the H$\beta$ emission line and to construct a map of the H$\alpha$/H$\beta$ ratio.  However, in the case where the dust is embedded in the ionized gas, scattered nebular light complicates the interpretation of the H$\alpha$/H$\beta$ ratio \citep{m70,s99}.  Instead, we chose to use the reddening values derived for the stars themselves to construct a reddening map.

We produced the reddening map through linear interpolation of the discrete reddening values.  We first constructed a mesh of triangles connecting nearby stars using Delaunay triangulation (through the IDL procedure TRIANGULATE), tiling the surface bounded by the locations of the observed stars.  We then linearly interpolated the reddening map over the grid using the values at the vertices.  Figure \ref{fig10} shows our H$\alpha$ image overlaid with smoothed contours of the reddening map.  The map shows that the brightest regions of DEM 228 tend to have higher reddening, but that the northern portion of the nebula is more heavily obscured than the southern end.
The peak reddening value is $E(B-V)=0.23$ while we assume $E(B-V)=0.04$ for the background reddening.  We used the reddening curve published by \citet{m90} to correct the H$\alpha$ surface flux for absorption by dust.

\subsection{Comparison of H$\alpha$ flux with stellar ionizing radiation}
Is the H$\alpha$ emission from DEM 228 fully accounted for by LH 72's O and B stars and those of the field?  We estimated the luminosity of Lyman continuum photons $Q_\circ$ produced by the observed hot stars using the calibration of \citet{sdk97}.  This calibration is based on theoretical spectral energy distributions of massive stars, calculated using models incorporating non-LTE effects, stellar winds, and line blanketing.  The parameters affecting $Q_\circ$  are temperature, gravity, and metallicity.  To derive the gravities, we estimated the masses by fitting to the $Z=0.008$ Geneva tracks \citep{s93} and calculated the radii through $L=4\pi R^2\sigma T_{\rm eff}^4$, $L$ and $T_{\rm eff}$ being known through our spectra and photometry.  Next, we identified all entries in Schaerer \& de Koter's (1997) Table 3 with temperatures within the errors of the temperatures of the observed stars; from that group of entries, we chose the entry with $\log g$ closest to the calculated $\log g$.  Finally, we interpolated $Q_\circ$ for the metallicity of the LMC from the $Z=0.02$ and $Z=0.004$ entries.

Assuming Case B recombination \citep{o89} and an LH 72 distance modulus of $(m-M)_\circ=18.4$, we next derived the H$\alpha$ fluxes that would be produced by the computed Lyman continuum luminosities.  To get an idea of the contribution of each star to the morphology of the \ion{H}{2} region, we then identified those circular regions in the de-reddened H$\alpha$ image which can account for all of the stellar ionizing radiation, taking into account all stars simultaneously.  Figure \ref{fig11} shows the de-reddened H$\alpha$ image with the circular regions overlaid.  The two most luminous stars, S128 and S133, are marked with X's.  Their Lyman continuum photons account for more H$\alpha$ emission than is measured in the image;  DEM 228 thus appears to be a density-bounded \ion{H}{2} region.  The remaining stars account for the bulk of the high-surface brightness emission.  Most of this emission appears to derive from multiple sources; for example, at least 15 sources contribute to the ionization of the southernmost complex of \ion{H}{2} regions.  On the other hand, two of the bright knots appear to be single-star \ion{H}{2} regions.

Although the total reddening-corrected H$\alpha$ emission accounts for only 60\% of the Lyman continuum photons produced, not all of the structure in the ionized gas is readily explained by the distribution of hot stars.  In particular, the low surface brightness halo of H$\alpha$ emission contains structure suggesting past star formation activity.  If true, this implies that the area hosting star formation within LH 72 has shrunk somewhat and drifted northwards.

\section{HI structure and kinematics}
Figure \ref{fig12} shows the individual channel maps of the HI cube covering the velocity range $272.5<v<320.3$ km s$^{-1}$.  The eighth-note shape of DEM 228 is mimicked by the nearby HI gas, most prominently between the velocities $300\lesssim v\lesssim310$ km s$^{-1}$.  In \citet{k99}, we identified this gas as forming the western  edge of a supergiant shell, SGS-14, an observation also made by \citet{ee98}.  However, because of the apparent lack of evidence for an approaching or receding cap to the shell, we did not then report an expansion velocity for it.

We now propose that the surface of SGS-14 is not smooth but patchy, and that the HI data indeed suggest that the shell is expanding.  We base this claim mainly on the evolution of the HI gas associated with DEM 228 through the sequence of channel maps shown in Figure \ref{fig12}.  In the $v=300.5$ km s$^{-1}$ map, an outline similar in shape to DEM 228 is clearly seen.  Following the northeastern tip of the outline, we notice that the tip moves progressively towards the interior of the shell with increasing velocity.  The first position-velocity diagram shown in Figure \ref{fig13} also demonstrates this behavior of the gas, indicating that the filament associated with DEM 228 is expanding with a velocity of $\sim15$ km s$^{-1}$.

The channel maps showing the eastern rim of SGS-14 also suggest expansion.  In the $v=302$ km s$^{-1}$ map, the faint trace of the ``C''-shaped eastern shell edge can be seen.  The arms of the ``C'' pinch together with increasing velocity, as would be expected if the gas formed part of an expanding shell.  The ``C'' fades into the noise at $v\sim315-320$ km s$^{-1}$, roughly the same velocity at which the tip of the filament forming the western edge of SGS-14 disappears.  The second position-velocity diagram shown in Figure \ref{fig13} implies an expansion velocity of $\sim$10 km s$^{-1}$.

The $v\sim274-284$ km s$^{-1}$ maps contain what may be a piece of the approaching portion of SGS-14.  The elongated cloud of gas lies roughly halfway between the geometric center of SGS-14 and its incomplete southern boundary.  The mean velocity of the cloud, $v\sim280$ km s$^{-1}$, suggests an expansion velocity of $\sim$20 km s$^{-1}$, as shown in the third panel of Figure \ref{fig13}.

The mass of HI that is involved in the 10-20 km s$^{-1}$ expansion can be calculated from the observed column density of the gas.  The lower left panel of Figure \ref{fig0} shows the assumed physical boundary of SGS-14.  Integrating over this area and the velocity range (280,320) km s$^{-1}$, we find a mass $M_{\rm SGS-14} = 1.8\times10^5$ M$_\odot$, with 3.3$\times10^4$M$_\odot$ purely in the gas associated with LH 72.  The kinetic energy carried by the \ion{H}{1} of SGS-14 is thus $\sim4\times10^{50}$ ergs, or that produced by $\sim$4 supernovae if one assumes that 10\% of the explosion energy ($\sim10^{51}$ ergs) goes into moving the surrounding ISM \citep{tt91}.

Assuming a distance modulus $(m-M) = 18.4$, the radius of SGS-14 is $\sim$240 pc.  Thus, if the gas in SGS-14 were uniformly distributed over the sphere, it would have an average density $\rho_\circ\sim0.1$ cm$^{-3}$.  
If SGS-14 expanded with constant velocity of $\sim$15 km s$^{-1}$ over its entire lifetime, its age is $\sim$16 Myr.

\section{Discussion and conclusion}
Having examined the stellar content of LH 72, studied the properties of the ionized gas of the embedded \ion{H}{2} region DEM 228, and derived the mass and kinematics of the surrounding \ion{H}{1} supershell SGS-14, a plausible story emerges: SGS-14 is an expanding supershell which has cleared the NE quadrant of LMC-4 of gas and triggered the formation of the OB association LH 72 behind its shock front.

What is the evidence?  First, the ages of the oldest stars in LH 72, $\sim15-30$ Myr, agree closely with the expansion age of SGS-14, $\sim$15 Myr.  Since the first triggering by the passing supershell, star formation could then have continued in LH 72 until $\lesssim5$ Myr ago, although without the clear N-S spatial pattern reported by \citet{o97}.  Second, the interior of SGS-14 appears empty of gas; the \ion{H}{1} maps show that the 21-cm emission is at the level of the noise.  Moreover, DEM 228 is density-bounded, allowing ionizing photons from the UV-bright stars S128 and S133 to escape into the interior of SGS-14.  Third, overlaying our H$\alpha$ image with contours of \ion{H}{1} column density show that while DEM 228 and LH 72 lie close to the rim of SGS-14, they are displaced towards its {\em interior}.  Figure \ref{fig15} shows the relationship.  The plate solution for the H$\alpha$ image was calculated using stars in LH 72 with coordinates from SIMBAD.  The calculated positions of four of the brightest stars agree to within $\lesssim1\arcsec$ with those listed in the Tycho-2 catalog \citep{h00}; the astrometric solution should thus be excellent.  The HI astrometric solution is accurate to $<<1\arcmin$.  The figure shows that the peak H$\alpha$ emission is offset by 1-2 arminutes (15-30 pc) from the peak HI column density, towards the interior of SGS-14.  

This picture of the formation of SGS-14 and LH 72 has an interesting consequence for our understanding of the supergiant shell LMC-4.  LMC-4 has often been thought to have formed as a unit (e.g. Westerlund \& Mathewson 1966; Dopita et al. 1985; Efremov et al. 1998).  However, this work shows that one quadrant of LMC-4 consists of an independently expanding shell, which may have triggered the formation of at least one of LMC-4's OB associations, LH 72, along its rim.  Thus, a possible formation scenario of LMC-4 is one in which many smaller shells, some triggering star formation from the swept-up gas, have overlapped to produce the ring of \ion{H}{1} and \ion{H}{2} regions we see today.  In this scenario, the energy required to produce LMC-4 is greatly reduced from that needed if it formed as a single unit--the reason being that once a supershell blows out, it becomes difficult to impart additional energy to expansion within the disk plane (e.g. Mac Low \& McCray 1988).  LMC-4, if it formed as a unit, must have blown out long ago, as its diameter of $\sim$1.4 kpc is many times the scale height of the LMC's \ion{H}{1} disk of 180 pc \citep{k99}.  By contrast, with a radius of $\sim240$ pc, SGS-14 should be only in the early stages of blowout.

The scenario proposed here naturally merges with that of \citet{ee98}, in which Quadrant and Sextant formed from swept-up shells which have since merged with SGS-11.  Indeed, LH 72 may be a younger, less massive version of Quadrant and Sextant.  Like those two structures, it is arc-shaped, with an approximate center of curvature near the center of SGS-14.  In $\sim$15 Myr, by the time that the last of the current generation of ionizing stars have disappeared, SGS-14 will have approximately doubled in size, assuming constant expansion velocity.  Assuming that LH 72 also grows only in the azimuthal direction, it will still have an overdensity of stars at the tip of the main sequence compared to the field, by a factor $\sim$4.  LH 72 may then be classified as a stellar arc, shorter and less tightly curved than the current Quadrant.

\acknowledgments
We are grateful to the NOAO Time Allocation Committee for the time awarded to this project, and to the CTIO mountain staff for their excellent support.  We thank Bruce Elmegreen for enlightening discussions, and Paul Hodge for the ideas that led to this project.  We thank the anonymous referee for prompt and helpful comments.





\clearpage
\figcaption[fig1.eps]{ \label{fig0}}
A montage showing the relationships of LH 72 and DEM 228 to SGS-14, LMC-4, and the LMC.  Proceeding counter-clockwise from top left, the images are: Boyden Observatory photograph of the LMC, courtesy of Paul Hodge and Harlow Shapley; same photograph with close-up of region containing LMC-4; section of \ion{H}{1} map \citep{k98} showing the location of SGS-14; true-color $UBV$ combined image of OB association LH 72 taken with the CTIO 1.5-m telescope; CTIO 1.5-m H$\alpha$ image of the \ion{H}{2} region DEM 228. 

\figcaption[fig2a.eps,fig2b.eps]{$(a)$ continuum-flattened, wavelength-calibrated spectra of 9 O and B stars in LH 72.  All spectra are plotted on the same arbitrary flux scale.  The locations of a number of typical absorption lines are indicated by vertical lines.  $(b)$ as in $a$, but showing stars containing Balmer emission lines. \label{fig1}}

\figcaption[fig3.eps]{ \label{fig2}}
Comparison of our $UBV$ photometry with that from the Magellanic Clouds Photometric Survey \citep{zht}.  Points with positive values along the $y$ axes indicate stars for which our photometry is fainter than that of the MCPS.  The solid lines are medians through the points, while the dotted lines are 1$\sigma$ deviations of the means.
 
\figcaption[fig4.eps]{ \label{fig3}}
Comparison of $U-B$ colors derived from our photometry, after correction for reddening as described in the text, with the intrinsic $U-B$ colors derived from our spectral classifications.  For most of the stars, the colors derived from the spectra are bluer than those found from the photometry, although the median difference is only $\sim$0.03 magnitudes.

\figcaption[fig5a.eps,fig5b.eps]{ \label{fig4}}
$(a)$ $B-V,V$ color-magnitude diagram of LH 72 and its surrounding field.  Error bars, shown as a function of $V$ magnitude, are the median $1\sigma$ photometric errors reported by DAOPHOT.  $(b)$ Color-magnitude diagram containing stars found only within the borders of LH 72, as defined by \citet{l72}.  Small points are stars removed by the subtraction procedure, while large circles are those that remain (see text for details).

\figcaption[fig6.eps]{ \label{fig5}}
H$\alpha$ magnitudes, defined as $-2.5\log(f_{H\alpha})+25$, plotted versus $B-V$; $f_{H\alpha}$ is measured in units of 10$^{-16}$ ergs cm$^{-2}$ s$^{-1}$.  Stars with spectra are labelled with filled circles.  Those stars with H$\alpha$ fluxes greater than or equal to those of LH72-1, L10, and S128 and with $B-V<0.5$ are considered to be candidate Be stars.

\figcaption[fig7.eps]{ \label{fig6}}
LH 72 color-magnitude diagram showing the locations of the candidate Be stars (filled circles) selected through Figure \ref{fig5}.  The Be candidates have red colors compared to the main sequence.

\figcaption[fig8.eps]{ \label{fig7}}
Histogram of reddening values $E(B-V)$ for stars within LH 72.  The distribution has a median of $E(B-V)=0.075$, mean of 0.09, standard deviation of 0.055, and a sharp lower bound at $(B-V)=0.04$.  The typical error on an individual measurement is $0\fm01$.

\figcaption[fig9.eps]{ \label{fig8}}
Hertzprung-Russell diagram of stars in LH 72, excluding candidate Be stars.  Black circles indicate stars with both spectroscopy and photometry, gray circles stars with photometry only.  The grey lines are evolutionary tracks from \citet{s93}, with initial masses labelled to the left of the tracks.  The black lines are isochrones with ages of 1, 2, 4, 8, 16, and 32 Myr.

\figcaption[fig10.eps]{ \label{fig9}}
H-R diagram produced if the observed $Q$ values are lowered by 0.05 magnitudes.  Evolutionary tracks and isochrones are as in Figure \ref{fig8}.

\figcaption[fig11.eps]{ \label{fig10}}
H$\alpha$ image overlaid with smoothed contours of the reddening map, which was derived by interpolating over stellar $E(B-V)$ values as described in the text.  The contours are labelled by $E(B-V)$; lines perpendicular to the contours indicate the direction of the local downhill gradient.

\figcaption[fig12.eps]{ \label{fig11}}
H$\alpha$ image, stretched by the square root of the intensity, overlaid with circles indicating the regions which account for all of the Lyman continuum emission from the enclosed hot stars (see text for an explanation of the calculation).  The stars S128 and S133 are marked by X's--their Lyman continuum contributions exceed the available H$\alpha$ flux.  The observed H$\alpha$ emission accounts for only 60\% of the available Lyman continuum photons, indicating that DEM 228 is density-bounded.

\figcaption[fig13.eps]{ \label{fig12}}
Individual \ion{H}{1} channel maps of the region surrounding SGS-14, spaced by units of the velocity resolution.  Velocities are in km s$^{-1}$; the spatial scale is indicated in the first channel.  The gas associated with LH 72/DEM 228 is most prominent in the velocity range $295.5 \le v \le 313.7$ km s$^{-1}$.

\figcaption[fig14.eps]{ \label{fig13}}
Position-velocity maps at three positions.  In each panel, the \ion{H}{1} map of SGS-14 is shown, summed over a range of velocities.  The white lines simulate a slit--the velocity structure at that slit position is shown either above or to the right of the spatial map.  Velocity labels are in km s$^{-1}$.

\figcaption[fig15.eps]{ \label{fig15}}
H$\alpha$ image, stretched by the logarithm of the intensity, overlaid with contours of \ion{H}{1} column density in units of $10^{21}$ atoms cm$^{-2}$.  North is up and east is to the left.

\clearpage
\begin{deluxetable}{lccccl}
\tabletypesize{\scriptsize}
\tablecaption{Spectral and Luminosity Classes of Stars in LH 72 \label{tbl-1}}
\tablewidth{0pt}
\tablehead{
\colhead{Star} &
\colhead{R.A. (2000.0)} &
\colhead{Dec.} &
\colhead{Sp. type} &
\colhead{Lum. class}  & 
\colhead{Comment} 
}
\startdata
S133   & 5$^{\rm h}$ 32$^{\rm m}$ 29\fs2     &  -66$^\circ$ 28$^\prime$ 00\farcs2 & O6 & V & \\
LH72-1 & 5$^{\rm h}$ 32$^{\rm m}$ 27\fs8   &    -66$^\circ$ 28$^\prime$ 20\farcs4 & O8.5 & V & \\
L29    & 5$^{\rm h}$ 32$^{\rm m}$ 19\fs9      & -66$^\circ$ 28$^\prime$ 02\farcs9  & O9 & III & \\
L41    & 5$^{\rm h}$ 32$^{\rm m}$ 18\fs4      & -66$^\circ$ 25$^\prime$ 51\farcs1  & O9.5 & V & \\
S128   & 5$^{\rm h}$ 31m 53\fs4     &  -66$^\circ$ 31$^\prime$ 14\farcs3  & O9.7 & Ib & \\
L22    & 5$^{\rm h}$ 32$^{\rm m}$ 18\fs1     &  -66$^\circ$ 28$^\prime$ 47\farcs2 & B0 & V & \\
L49    & 5$^{\rm h}$ 32$^{\rm m}$ 13\fs0      & -66$^\circ$ 25$^\prime$ 03\farcs0 & B0.2 & V & \\
L51    & 5$^{\rm h}$ 32$^{\rm m}$ 20\fs4      & -66$^\circ$ 23$^\prime$ 58\farcs2  & B0.5 & III & \\
S135   & 5$^{\rm h}$ 32$^{\rm m}$ 33\fs2      & -66$^\circ$ 26$^\prime$ 12\farcs3   & B3 & Ia-b & \\
L10    & 5$^{\rm h}$ 32$^{\rm m}$ 33\fs7      & -66$^\circ$ 27$^\prime$ 10\farcs0  & O8 & V & emission lines\\
L39    & 5$^{\rm h}$ 32$^{\rm m}$ 09\fs9     &  -66$^\circ$ 26$^\prime$ 08\farcs4 & B0.5 & V & emission lines \\
S132   & 5$^{\rm h}$ 32$^{\rm m}$ 18\fs8     &  -66$^\circ$ 24$^\prime$ 12\farcs0  & B8 & Ia & emission lines \\
\enddata




\end{deluxetable}

\begin{deluxetable}{lccccc}
\tablecaption{Photometric transformation coefficients \label{tbl-2}}
\tablewidth{0pt}
\tablehead{
\colhead{Date} &
\colhead{$A_0$} & \colhead{$A_1$} & \colhead{$A_2$} & \colhead{$A_3$} & \colhead{$A_4$}
}
\startdata
11/25/98  & 4.249(7)  & -0.167  & 0.136  & 0.374(60) & 0.0 \\
11/26     & 4.323(5)  & -0.167  & 0.136  & 0.555(42) & -0.009(2) \\
11/27     & 4.278(5)  & -0.167  & 0.136  & 0.485(25) & 0.0 \\
\tableline
 &  $B_0$ &  $B_1$ &  $B_3$ &  $B_4$ \\
\tableline
11/25/98  & 2.414(3)  & 0.112  & 0.287(30)  & 0.0  \\
11/26     & 2.426(2)  & 0.112  & 0.302(16)  & -0.003(1)  \\
11/27     & 2.391(2)  & 0.112  & 0.260(10)  & 0.0    \\
\tableline
 &  $C_0$ &  $C_1$ &  $C_3$ &  $C_4$ \\
\tableline
11/25/98 & 2.084(2)  & -0.029  & 0.149(18)  & 0.0 \\
11/26   & 2.102(2)  & -0.029  & 0.139(14)  & -0.005(1) \\
11/27   & 2.091(1)  & -0.029  & 0.140(5)   & 0.0 \\
\enddata
\end{deluxetable}

\end{document}